\documentclass[twocolumn,showpacs,preprintnumbers,amsmath,amssymb]{revtex4}

\usepackage{graphicx}
\usepackage{dcolumn}
\usepackage{bm}
\usepackage[latin1]{inputenc}

\begin{document}

\title{Entangled electronic state via an interacting quantum dot}

\author{Gladys Le\'on$^{1,2}$, Otto Rendon$^{1,3}$, and Ernesto Medina$^{1,*}$}
\affiliation{$^1$Centro de F\'\i sica, Instituto Venezolano de Investigaciones Cient\'{\i }ficas. 
Apartado 21827, Caracas 1020 A, Venezuela.}
\affiliation{$^2$Departamento de F\'{\i}sica-Facultad de Ciencias, Universidad Central de 
Venezuela, Caracas, Venezuela.}
\affiliation{$^3$Departamento de F\'{\i}sica-FACYT, Universidad de
Carabobo, Valencia, Venezuela.}
\affiliation{*Corresponding author (ernesto@pion.ivic.ve)}
\author{Horacio M. Pastawski$^4$, Vladimiro Mujica$^5$}
\affiliation{$^4$Facultad de Matem\'atica, Astronom\'{i}a y F\'{i}sica, LaNAIS de RMS 
CONICET, Universidad Nacional de C\'ordoba, Ciudad Universitaria, 
5000 C\'ordoba, Argentina}
\affiliation{$^5$Departamento de Qu\'{\i}mica-Facultad de Ciencias, Universidad Central de 
Venezuela, Caracas, Venezuela.}

\date{\today}

\begin{abstract}
We study a device for entangling electrons as cotunneling occurs through a
quantum dot where on-site electron-electron interactions $U$ are in place. The main advantage of this device is that single particle processes are forbidden by energy conservation as proposed by Oliver et al\cite{oli02}. Within this model we calculated two electron transition amplitude, in terms of the T-matrix, to all orders in the coupling to the dot, and consider a finite lead bandwidth.  The model filters singlet entangled pairs with the sole requirement of Pauli principle. Feynman paths involving consecutive and doubly occupied dot interfere destructively and produce a transition amplitude minimum at a critical value of the onsite repulsion $U$. Singlet filtering is demonstrated as a function of a gate voltage applied to the dot with a special resonance condition when the dot levels are symmetrically placed about the input lead energy.
\end{abstract}

\pacs{03.65.Ud,73.63.Kv}
\keywords{}
\maketitle

In the last few years, a number of methods have been proposed for generating entanglement between spin states of electrons\cite{loss98,bur99,barn00,rech01,costa01,oli02}. Such Einstein-Podolsky-Rosen states are a vital resource in securing quantum communication, teleportation\cite{ben00}, and cryptography\cite{eke91}. Furthermore, long spin decoherence times in semiconductors\cite{fuji01}, approaching a few microseconds and the concomitant phase coherent transport over distances of the order of $0.01$ cm, make entangled electron spins a good candidate for quantum computing applications.

\begin{figure}[htbp]
\includegraphics[width=3.5in]{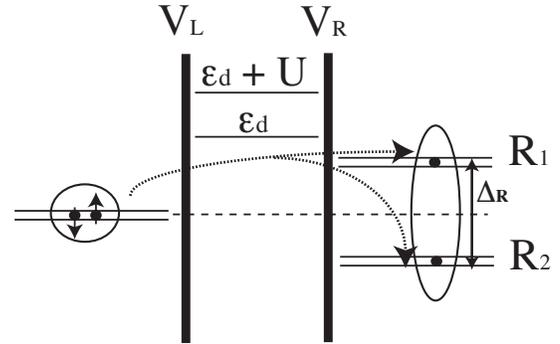}
\caption{\label{figura1} Energy level diagram of the model. The external leads are coupled to the dot with a coupling strength $V_{L,R}$. Electron-electron interactions are only considered within the dot. The energy difference between the outgoing leads is $\Delta_R$, and they are positioned symmetrically about the input electron energy.}
\end{figure}

We have focused on a very simple model entangler proposed by Oliver {\it{et al.}} 
\cite{oli02}, because of the convenient suppression of single-electron tunneling
by energy conservation. Such a model brings about important issues of the role of electron-electron interactions in the dot and the effects of coupling to the leads observed in continuum treatments of electron-electron interactions\cite{ent00}. The model consists of one input and two output leads attached to a quantum dot with no occupied states (Fig.~\ref{figura1}). The arrangement of levels is such that single or double occupancy of the dot does not conserve energy and thus only virtual states can comply within the energy uncertainty. A virtual double occupancy of the dot incurs in an on-site Coulomb energy $U$. The external contacts are considered non-degenerate leads, with a relatively narrow but finite energy bandwidth, where the independent electron approximation is still valid. Single electron transition amplitude is avoided by placing the incoming and the two outgoing leads off resonance. However, the lead energies can be arranged so that two-electron co-tunneling events conserve energy (see Fig.~\ref{figura1}).

The work of Oliver {\it {et al}} analyzed the zero input/output bandwidth limit  of the previous model, thus neglecting dot level broadening. They found that while the transition amplitude of the triplet state is always zero, the singlet state has a finite transition probability only possible due to a finite electron-electron interaction energy, i.e. entanglement (singlet production) is mediated by interaction. Nevertheless the limit considered in ref.\cite{oli02} ignores the coupling of the system to leads and possible sources of dephasing effects (inelastic or energy conserving\cite{bonca} ) on the filtering properties of the interacting dot, which are necessarily present in experimental situations.  Here we discuss two important issues regarding the transition amplitude in this model: a) the effect of broadening  due to coupling to finite bandwidth semi-infinite leads on the mediating role of the on site electron-electron interaction and the production of the singlet entangled states, and b) the behavior of the model in resonance conditions between the internal dot states and the external leads to study the robustness of the filtering property. 

This paper is organized as follows: we first present the model and the calculational methodology used for semi-infinite leads as input/outputs terminals. We then discuss the topology that connects the different two particle states of the system and the state basis in which the system is to be diagonalized. The results show the filtering of singlet pairs  as a function of the coupling to the dot. Following, we include semi-infinite leads through complex self energies coupled to the dot and find qualitatively different results from the case of zero bandwidth finding that Pauli principle alone is sufficient to secure generation of entangled states. We end with a discussion contemplating the transition amplitude as a function of an external gate voltage on the dot and the special resonance conditions found. 

The system is described by a tight-binding Hamiltonian with an on-site Coulomb energy term $U$ with no single-electron excitations within the dot
\begin{eqnarray}\label{hamiltoniano}
\hat{\mathcal{H}}&=&\hat{\mathcal{H}}_{\mathrm{0}}+\hat{{V}}\cr
\hat{\mathcal{H}}_{\mathrm{0}} &=&\sum_{s,k,\sigma} \varepsilon_{s,k}
\hat{a}^\dagger_{s,k,\sigma}\hat{a}_{s,k,\sigma}+\sum_{\sigma}\varepsilon_{d}
\hat{c}^\dagger_{\sigma}\hat{c}_{\sigma}+{U}\hat{n}_{\uparrow}
\hat{n}_{\downarrow}\cr\hat{V}&=&\sum_{s,k,\sigma}\left(V_{s}
\hat{a}^\dagger_{s,k,\sigma}\hat{c}_{\sigma}+
\mathrm{c.c.}\right)
\end{eqnarray}
The first term in ${\hat{\mathcal H}_{0}}$ describes the leads. The operator
${\hat{a}^\dagger_{s,k,\sigma}}$ (${\hat{a}_{s,k,\sigma}}$)
creates (annihilates) an electron of momentum $k$ and spin $\sigma$ in
either lead according to the lead label, where $s = \left\{L,R_1,R_2\right\}$, for left and right $1$ and $2$ leads respectively, and $\sigma =\{\uparrow,\downarrow\}$. We will also consider the case of localized external states of quantum number $k,k'$. The second and third term of ${\hat{\mathcal H}_{0}}$ describe the isolated quantum dot with an intra-dot interaction $U$. The operator ${\hat{c}^\dagger_{\sigma}}$ ( ${\hat{c}_{\sigma}}$ ) creates (annihilates) an electron with single energy level degenerate in spin. The coupling term ${\hat{V}}$ is the off diagonal part of the Hamiltonian that characterizes the transfer of electrons between the leads and dot.  As in reference\cite{oli02} we take the initial energy $E_i=\varepsilon_{L,k}+\varepsilon_{L,k'}$ and the final energy $E_f=\varepsilon_{R_1,k_1}+\varepsilon_{R_2,k_2}$ where $\varepsilon_{L,k}\neq\varepsilon_{R_1,k_1}\neq\varepsilon_{R_2,k_2}$, eliminating single electron tunneling, and we denote the two electron energy with the capital $E$. Two electron tunneling can occur when $E_i=E_f$ so that each electron ends up on different outgoing leads.  If we define $\Delta_L=\frac{1}{2}(\varepsilon_{L,k}-\varepsilon_{L,k'})$ and $\Delta_R=\frac{1}{2}(\varepsilon_{R_1,k_1}-\varepsilon_{R_2,k_2})$ the outgoing states have energy $\varepsilon_{R_1,k_1},\varepsilon_{R_2,k_2}=\frac{1}{2}E_i\pm\Delta_R$.
where $\Delta_L<\Delta_R$ for single electron tunneling to be suppressed. This situation is considered throughout this paper. As a simplifying choice all $V_{s}=V$ and the lower level in the dot is $\varepsilon_d=0$, and serves as the energy reference of energy for the system.

\begin{figure}[b]
\includegraphics[width=3.5in]{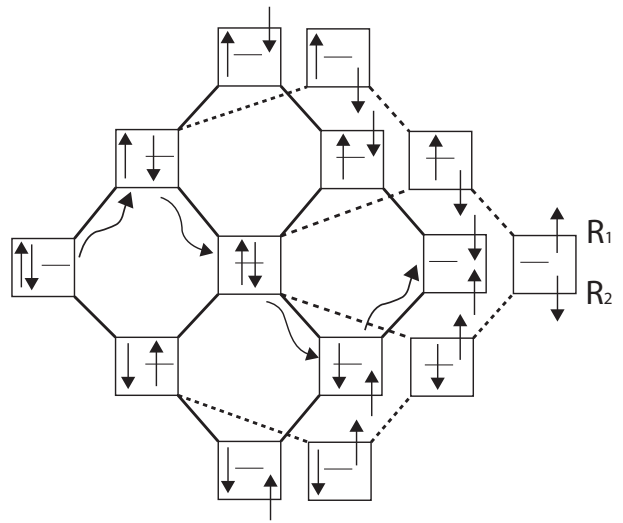}
\caption{\label{figura2} Lattice of states connecting the initial and final spin configurations. The segment within the box denotes the quantum dot. When the arrows appear to the right at the upper and lower sides of the box they denote spins on the upper $R_1$ and lower $R_2$ outgoing channels. The wavy lines denote one possible directed Feynman path contributing to fourth order perturbation theory of reference \cite{oli02}. }
\end{figure}

The T-matrix formalism is used to compute the {\it transition amplitude} between the initial and final states $\left|i\right >$ and $\left|f\right >$ using the recursive expression
\begin{equation}\label{tmatrixrecursive}
\hat{T}(E)=\hat{V}+\hat{V}\frac{1}{E-\hat{{H}_{0}}}
\hat{T}(E).
\end{equation}
In the limit of small lead-dot coupling and zero lead band-width limit, Oliver {\it{et al.}} calculated the $T$-matrix through a perturbation expansion in the tunneling matrix element $V$ to lowest order for the transition amplitude. They considered an initial two-electron state, $\left|{i}\right> =
\hat{a}^{\dagger}_{s,k,\sigma}\hat{a}_{s,\kappa,\sigma}
\left|{0}\right>$ and a final state $\left|{f}\right>$ that can be either a singlet
$\left|{s}\right> =
\frac{1}{\sqrt{2}}\left(\hat{a}^{\dagger}_{R_{1}\uparrow}
\hat{a}^{\dagger}_{R_{2}\downarrow}-\hat{a}^{\dagger}_{R_{1}\downarrow}
\hat{a}^{\dagger}_{R_{2}\uparrow} \right)\left|{0}\right> $ or a triplet
$\left|{t}\right> =
\frac{1}{\sqrt{2}}\left(\hat{a}^{\dagger}_{R_{1}\uparrow}
\hat{a}^{\dagger}_{R_{2}\downarrow}+\hat{a}^{\dagger}_{R_{1}\downarrow}
\hat{a}^{\dagger}_{R_{2}\uparrow} \right)\left|{0}\right> $, with
$\left|{0}\right>$ the vacuum. Fourth order perturbation predicts that the entangler device filters the singlet portion of the initial state $\left|{i}\right> =
\hat{a}^{\dagger}_{s,k,\sigma}\hat{a}_{s,k,\sigma}
\left|{0}\right>$ (input) and generates a non-local spin-singlet state
at the output leads. The singlet state vanishes when the intra-dot Coulomb
interaction is turned off. Thus,  the Coulomb interaction acts as the mediator of entanglement. Note that although the initial state is correlated by indistinguishability, it can be represented by a single Slater determinant while the final state $\left | f\right >$ is a sum of products of Slater determinants. In the limit of large separations, electrons are effectively distinguishable entities. 

We go beyond the previous approach by including finite bandwidth to terminals attached to the dot and also check all computations to all orders in perturbation theory by fully diagonalizing in a chosen basis. To introduce the input/output leads we add an appropriate self-energy $\Sigma_{\eta}$ \cite{pas01,pals96} to external sites coupled to the dot. The semi-infinite leads have a semi-circular density of states and a band width of $4V_{lead}$. The transition amplitude of pairs of electrons is calculated exactly between a Slater determinant input state $\left|{i}\right>$ and a non-local, singlet or triplet  output state $\left|{f}\right>$ defined previously.  

Based on the tight-binding Hamiltonian of Eq.\ref{hamiltoniano}, 
one can build the diagram shown in Fig.\ref{figura2}. Each box in the figure shows either real (initial and final) or virtual (intermediate) states of two electrons, with their spins. The horizontal line in each box represents the quantum dot. The entering state is on the left and the outgoing states are on arms $R_1$ or $R_2$ indicated by placing the spins on the upper and lower sides of the boxes. When the electrons are in the dot the spin states are shown on the horizontal line within the box. All Feynman amplitudes are built by summing paths (directed or not) on this lattice of configurations. The wavy lines indicate one of the possible {\it directed path} of fourth order in $V$ considered in perturbation theory by Oliver et al\cite{oli02}.  The diagram then represents a Hamiltonian in a Hilbert space of 14 dimensions.

The dot is connected to open semi-infinite discrete leads that dress the single particle energies at external sites, with a complex self energy $\Sigma(\varepsilon)=\varepsilon/2-{\rm i}\sqrt{V_{lead}^2-(\varepsilon/2)^2}$, within  a band of width $4V_{lead}$, that shifts and broadens the internal dot levels. The input electrons have energy $E_i/2$ ($\Delta_L=0$) and the output bands are separated by an energy $\Delta_R$ measured from the center of the respective bands. The transition matrix $\mid\left<{f}\right|\hat{T}(E_{i}+{\rm i}\eta)\left|{i}\right>\mid$ is computed using the relation 
\begin{equation}\label{tmatrixgreen}
\hat{T}(E+{\rm i}\eta)=\hat{V}\hat{G}(E+{\rm i}\eta)\hat{G}_{0}^{-1}
(E+{\rm i}\eta)
\end{equation}
where $\hat{G}(E+{\rm i}\eta)$ is the Green's function of the full Hamiltonian
$\hat{\mathcal{H}}$ and
$\hat{G}_{0}^{-1}(E+{\rm i}\eta)$ is the Green's function of the unperturbed Hamiltonian $\hat{\mathcal{H}}_{\mathrm{0}}$ as defined above.
The Green's functions are computed by fully inverting $(E-\mathcal{H})$
in the basis defined in Fig. \ref{figura2}.

\begin{figure}[tb]
\includegraphics[width=3.2in]{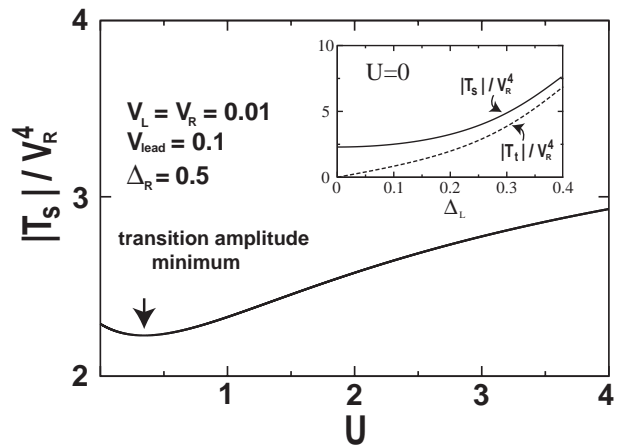}
\caption{\label{figura3} Singlet transition amplitude for the interacting dot coupled to external leads with the parameters given in terms of the input lead bandwidth. The exact results indicate singlet generation even in the absence of on site repulsion but complying with the Pauli exclusion principle. The existence of a singlet transition amplitude minimum is pointed out as a function of the interaction $U$. The triplet amplitude is zero to all orders for degenerate input electrons ($\Delta_L=0$). The inset shows the behavior of the triplet amplitude, as compared to the singlet, when the input degenracy is lifted. }
\end{figure}

\begin{figure}
\includegraphics[width=3in]{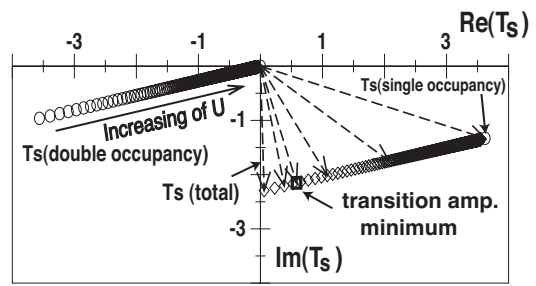}
\caption{\label{figura4} Complex transition amplitude $T_s$ for the same parameters of Fig. 3, separated in single and double dot occupancy components, computed exactly, as a function of the interaction energy $U$. The dotted line depicts the total sum of the two occupancy components. As $U$ increases the magnitude of the double occupancy component decreases making the total $T_s$ change in phase and magnitude and exhibiting a minimum (indicated by the square).}
\end{figure}

\begin{figure}
\includegraphics[width=2.7in]{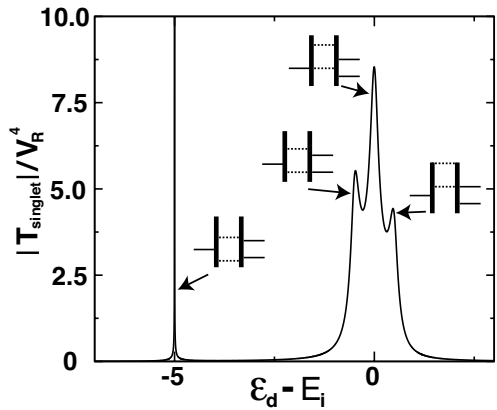}
\caption{\label{figura5} Singlet transition amplitude as a function of the energy difference between the input lead and the quantum dot single particle state. $U=10$, and the same lead and coupling parameters of Fig. 3. The resonances correspond to the coincidence between either the input lead or the output leads with the internal single particle state or when the two internal states of the dot are symmetrically positioned about the incoming energy.}
\end{figure}

Figure \ref{figura3} shows the singlet transition amplitude for the case where one dimensional leads are attached to the interacting dot as explained previously. For the range of parameters of interest the transition amplitude departs from a finite value at $U=0$ and develops a minimum at finite $U$. The transition amplitude then increases  towards a saturation value (at $U=\infty$), as the double occupancy terms become less important. These results indicate that entangled electrons can be produced only requiring the Pauli principle. Such processes are not allowed for zero bandwidth as in the case of ref.\cite{oli02}.   The triplet transition amplitude is zero to all orders of perturbation when the input electrons are degenerate. When the degeneracy is lifted i.e. $\Delta_L\neq 0$ a finite triplet amplitude ensues, as shown in the inset of Fig.\ref{figura3}. This implies a degradation of the entangler, since the limit in that $T_s/T_t\rightarrow 1$ no entanglement occurs. Vanishing of the triplet amplitude arises from the fact that virtual state energies, for both output lead exchanges, are the same leading to the same phase changes except for an overall minus sign coming from commutation relations. This symmetry is broken when input energies are non-degenerate. 

For small coupling to the leads, perturbation theory is sufficient to qualitatively analyze all the changes in the transition amplitude. Although the perturbative expressions are now very complex, it is easy to see that the double occupation terms (that depend on $U$) interfere destructively with the virtual consecutive tunneling contributions. The magnitude of the double occupation contributions monotonically decreases reaches a with the on-site repulsion towards zero. These effects are illustrated in Fig.\ref{figura4}, where the complex transition amplitude computed by fourth order perturbation is depicted as a function of the electron-electron interaction. The transition amplitude is separated into the double occupancy (involving interaction) and single occupancy amplitudes, which are added to give the total amplitude. Only the double occupation amplitude changes in magnitude (but not phase)  as a function of $U$, so that the total amplitude changes both in phase and magnitude. The total amplitude phasor moves along a straight line passing through a minimum magnitude (discussed in Fig.\ref{figura3}), a value depicted in the figure by a square.  Such mechanism precludes cancellation at $U=0$ in the presence of broadening. The same can be applied to the magnitude of the triplet, but now the interfering amplitudes are those coming from the exchange of output leads. In this case a symmetry between exchanged paths, which destroys the triplet when input electron are degenerate, is broken by a finite $\Delta_L$ (the energy difference of non-degenerate electrons\cite{future}).

The energy configuration of the external leads is the only ingredient necessary for the exclusion of single electron tunneling processes granted that no overlapping between input/output bands occurs (our case). Preserving this character of the system one can study the behavior of the transition amplitude as a gate voltage is applied to the dot changing the relative disposition of the energy levels. Figure \ref{figura5} depicts the transition amplitude for the singlet as a function of the difference in energy between the dots single particle state and the external input lead. The resonances correspond to the coincidence of the single particle dot level either with the outgoing (lateral peaks) or incoming leads (central peak). The third, very fine resonance, corresponds to the case where the internal dot levels are symmetrically placed around the incoming lead energy ($U/2+(\varepsilon_d-E_i)=0$).  The triplet amplitudes are zero in perturbation theory and to all orders in the coupling no matter what the gate voltage is. These results put less of a restriction to the value of the energy levels within the dot although it assumes only two levels.

In summary we have studied a model electron entangling device\cite{oli02} which precludes single electron processes. We have considered the coupling of the interacting dot to external leads with a finite bandwidth.  We find that, while triplet transition amplitude is always zero to all orders if the input electrons are energy degenerate, singlet entangled pairs have a finite transition amplitude even at on-site electron repulsion $U=0$. Thus, the Pauli principle seems sufficient for generating entangled electrons. On the other hand if the electrons at the input are non degenerate a finite triplet arises degrading the efficiency of the entangler as the energy difference increases. 

Are the parameters for operation of the entangler realistic? We can set the lead bandwidth,  for example using a series of quantum wells or coupled dots, at 30-50 meV (quantum well barrier height of 0.4eV and width of $\sim$30\AA of AlGaAs) so that the coupling to our interacting dot would have to be set in range of 3-5 meV with a broader inter-well barrier.  This would imply a transition rate of the order of 5300 (1/sec) for the singlet. This small transition rate is consistent with the co-tunelling process described. In spite of being small, such co-tunelling contributions to the current are dominant (disregarding temperature effects) because single particle events are forbidden since they do not conserve energy given that the input and output bands do not overlap.

Three resonance values for singlet transition amplitude are found as a function of the energy difference between the single particles dot level and the input lead; i) one for every coincidence between the single particle level and the input or put leads and ii) a non-trivial strong resonance for the case of symmetrically place dot levels about the input terminal energy. The doubly occupied paths interfere destructively with the virtual consecutive tunneling paths generating a transition amplitude minimum at a critical value of the doubly occupied dot energy.

\acknowledgments
We acknowledge financial support from CONICET (Argentina) and FONACIT (Venezuela). EM thanks J. Kinaret for useful discussions.

\end{document}